\documentclass[aps,prd,twocolumn,nofootinbib,showpacs]{revtex4}
\usepackage[dvips]{graphicx} \usepackage{amssymb} \usepackage{psfrag}
\usepackage{hyperref}\usepackage{amsmath}
\allowdisplaybreaks[2]
\newcommand{\eq}[1]{Eq. (\ref{#1})}
\newcommand{\fig}[1]{Fig. \ref{#1}}
\newcommand{\ocite}[1]{Ref. \cite{#1}}
\begin{document}
\begin{flushright}
\vspace*{-1.5cm}
BA-08-11\\
\end{flushright}
\title{Chaotic inflation, radiative corrections and precision cosmology}
\author{V. Nefer \c{S}eno\u{g}uz}   \email{nsenoguz@dogus.edu.tr}
\affiliation{Division of Sciences, Do\u{g}u\c{s} University,
34722 Kad{\i}k\"oy, Istanbul, Turkey}
\author{Qaisar Shafi} \email{shafi@bartol.udel.edu}
\affiliation{Bartol Research Institute, Department of Physics and Astronomy, University of Delaware, Newark, DE
19716, USA}
\begin{abstract}
We employ chaotic ($\phi^2$ and $\phi^4$) inflation to illustrate the important
role radiative corrections can play during the inflationary phase.
Yukawa interactions of $\phi$, in particular, lead to corrections of the form
$-\kappa\phi^4 \ln (\phi/\mu)$, where $\kappa >0$ and $\mu$ is a renormalization scale.
For instance, $\phi^4$ chaotic inflation with radiative corrections
looks compatible with the most recent WMAP (5 year) analysis,
in sharp contrast to the tree level case. We obtain the
 95\% confidence limits $2.4\times10^{-14}\lesssim\kappa\lesssim5.7\times10^{-14}$,
$0.931\lesssim n_s\lesssim0.958$ and $0.038\lesssim r\lesssim0.205$,
where $n_s$ and $r$ respectively denote the scalar spectral index
and scalar to tensor ratio. The limits for $\phi^2$ inflation are
$\kappa\lesssim7.7\times10^{-15}$,
$0.929\lesssim n_s\lesssim0.966$ and $0.023\lesssim r\lesssim0.135$.
The next round of precision experiments should
provide a more stringent test of realistic chaotic $\phi^2$ and $\phi^4$ inflation.
\end{abstract}
\maketitle

Chaotic inflation driven by scalar potentials of the type $V =
(1/2)m^2  \phi^2$ or $V = (1/4!)\lambda  \phi^4$ provide just about
the simplest realization of an inflationary scenario
\cite{Linde:1983gd}. For the $\phi^2$ potential, the predicted
scalar spectral index $n_s \approx 0.966$ and scalar to tensor ratio
$r\approx0.135$ are in good agreement with the most recent Wilkinson
Microwave Anisotropy Probe (WMAP) 5 year analysis
\cite{Dunkley:2008ie,Komatsu:2008hk}. For the $\phi^4$ potential,
the predictions for $n_s$ and $r$ lie outside the WMAP 95\%
confidence limits.

In this letter we wish to emphasize the fact that radiative
corrections can significantly modify the `tree' level predictions
listed above. The inflaton field $\phi$ must have couplings to
`matter' fields which allow it to make the transition to hot big
bang cosmology at the end of inflation. These couplings will induce
quantum corrections to $V$, which we take into account following the
analysis of Coleman and Weinberg \cite{Coleman:1973jx}.
(For a comparison of Coleman-Weinberg potential with WMAP, see \ocite{Shafi:2006cs}.)
Even if such terms are sub-dominant during inflation, they can make sizable
corrections to the tree level predictions for $n_s$ and $r$.

\enlargethispage{\baselineskip}

Here, we investigate the impact of quantum corrections on the simplest chaotic
($\phi^2$ and $\phi^4$) inflation models.
We do not consider a specific framework 
such as supergravity, where the potential generally gets modified and becomes
exponentially steep for
super-Planckian values of the field. 
(For a realization of chaotic inflation in supergravity, see \ocite{Kawasaki:2000yn}.)
We instead assume that quantum gravity
corrections to the potential become large only at super-Planckian
energy densities \cite{Linde:2005ht}, which can 
allow higher order terms to be negligible during
the observable part of inflation \cite{Linde:2007fr}.
We are mainly interested
in the coupling of $\phi$ to fermion fields, 
for these give rise to radiative corrections to $V$
which carry an overall negative sign. A simple example is provided
by the Yukawa coupling $(1/2)h \phi \bar{N} N$, where $N$ denotes
the right handed neutrino. (Note that $N$ may also have bare mass
terms.) Such couplings provide correction terms to $V$ which, to
leading order, take the form
\begin{equation} \label{deltav}
 V_{\rm loop}\approx -\kappa  \phi^4 \ln\left(\frac{h\phi}{\mu}\right)
\end{equation}
where $\kappa = h^4/(16\pi^2)$ in the one loop approximation, and
$\mu$ is a renormalization scale. The negative sign is a
characteristic feature for the contributions from fermions.

By taking into account the contribution provided by \eq{deltav}, we
find that depending on $\kappa$, the scalar to tensor ratio $r$ can be
considerably lower than its tree level value. An interesting consequence
is that $\phi^4$ inflation, which has been ruled out at tree level, becomes
viable for a narrow range of $\kappa$. The predictions for
$n_s$ and $r$ extend from the tree level values
to a new inflation regime of small $r$ and $n_s\ll1$. 
(A similar range of predictions can be obtained at tree level for the binomial
potential $V=V_0-(1/2)m^2\phi^2+(1/4!)\lambda  \phi^4$ \cite{Cirigliano:2004yh}.)
We can expect that the next
round of precision measurements of $n_s$, $r$ and related quantities such as
$\alpha\equiv{\rm d}n_s/{\rm d}\ln k$
will provide a stringent test of these more realistic $\phi^2$ and $\phi^4$ inflation models.

To see how the correction in \eq{deltav} arise, consider the Lagrangian density
\begin{equation}\label{lagrange}
\begin{split}
{\cal L} &=
\frac12\partial^{\mu}\phi_B\partial_{\mu}\phi_B+\frac{i}{2}\bar{N}\gamma^{\mu}\partial_{\mu}N
 -\frac12 m_B^2 \phi^2_B-\frac{\lambda_B}{4!} \phi_B^4\\ & \quad -\frac12 h \phi_B \bar{N} N
-\frac12 m_{N} \bar{N} N \,,
\end{split}
\end{equation}
where the subscript `B' denotes bare quantities, and the field $N$ 
denotes a Standard Model singlet fermion (such as a right
handed neutrino). The inflationary potential including one loop corrections is given by
\begin{equation}
V = \frac12 m^2 \phi^2 +\frac{\lambda}{4!} \phi^4+ V_{\rm loop}\,,
\end{equation}
where, following \ocite{Coleman:1973jx},
\begin{equation} \label{loop}
\begin{split}
V_{\rm loop} &= \frac{1}{64\pi^2} \Bigg[
\left(m^2+\frac{\lambda}{2}\phi^2\right)^2\ln\left(\frac{m^2+(\lambda/2)\phi^2}{\mu^2}\right)\\
& \quad
-2(h\phi+m_N)^4\ln\left(\frac{\left(h\phi+m_N\right)^2}{\mu^2}\right)
\Bigg]\,.
\end{split}
\end{equation}
For the range of $h$ that we consider, $h \phi\gg m$ and $h^2\gg\lambda$
during inflation. Also assuming $h \phi\gg m_N$,
the leading one loop quantum correction to
the inflationary potential is given by \eq{deltav}.
Note that with $h\phi\gg H$ (Hubble constant), the `flat space'
quantum correction is a good approximation during inflation. (For a discussion of
pure Yukawa interaction involving massless fermions in a locally de Sitter geometry
see \ocite{Miao:2006pn}. For a discussion of one-loop effects in chaotic
inflation without the Yukawa interaction see \ocite{Sloth:2006az}.)
For convenience, we will set the renormalization
scale $\mu=h m_P$, where $m_P\approx2.4\times10^{18}$ GeV is the (reduced)
Planck scale. (Changing the renormalization scale corresponds to redefining $\lambda$,
and does not affect the physics.)

The instability for $\phi\gg m_P$ caused by the negative
contribution of \eq{deltav} will not concern us too much here. Presumably it is taken care of
in a more fundamental theory. Our inflationary phase takes place for
$\phi$ values below the local maximum. Although this differs from the original
chaotic inflation model, it is still possible to justify the initial
conditions. Inflation most naturally starts at an energy density
close to the Planck scale. However, the observable part of inflation occurs at
a much lower energy density. If, after the initial phase of inflation, there
exist regions of space where the field is sufficiently close to the local
maximum, eternal inflation takes place. It would then seem that the regions
satisfying the condition for eternal inflation would always dominate, since
even if they are initially rare, their volume will increase indefinitely. For
discussions of this point, see e.g. Refs. \cite{Linde:2005ht,Vilenkin:1983xq,Boubekeur:2005zm}.

Before we discuss the effect of \eq{deltav} on the inflationary parameters,
let's recall the basic equations.
The slow-roll parameters may be defined as
(see \ocite{Liddle:2000cg}
for a review and references):
\begin{equation}
\epsilon =\frac{1}{2}\left( \frac{V^{\prime} }{V}\right) ^{2}\,, \quad
\eta = \frac{V^{\prime \prime} }{V}  \,, \quad
\xi ^{2} = \frac{V^{\prime} V^{\prime \prime\prime} }{V^{2}}\,.
\end{equation}
Here and below we use units $m_P=1$, and `$\prime$' denotes derivative with respect to $\phi$.
The spectral index
$n_s$, the tensor to scalar ratio
$r$ and the running of the spectral index
$\alpha\equiv\mathrm{d} n_s/\mathrm{d} \ln k$ are given by
\begin{eqnarray}
n_s \!&=&\! 1 - 6 \epsilon + 2 \eta \label{ns}\,,\\
r \!&=&\! 16 \epsilon \,,\\
\alpha \!\!&=&\!\!
16 \epsilon \eta - 24 \epsilon^2 - 2 \xi^2\,.
\end{eqnarray}
The amplitude of the curvature perturbation $\Delta_R$ is given by
\begin{equation} \label{perturb}
\Delta_R=\frac{1}{2\sqrt{3}\pi}\frac{V^{3/2}}{|V^{\prime}|}\,.
\end{equation}
The WMAP best fit value for
the comoving wavenumber $k_0=0.002$ Mpc$^{-1}$ is $\Delta_R=4.91\times10^{-5}$
\cite{Dunkley:2008ie}.

In the slow-roll approximation, the number of e-folds is given by
\begin{equation} \label{efold1}
N_0=\int^{\phi_0}_{\phi_e}\frac{V\rm{d}\phi}{V^{\prime}}\,, \end{equation}
where the subscript `0' implies that the values correspond to $k_0$. The subscript `e'
implies the end of inflation, where $\epsilon(\phi_e)\simeq1$.
$N_0$ corresponding to the same scale is \cite{Kolb:1990vq}
\begin{equation} \label{nuk}
N_0\approx65+\frac12\ln[V(\phi_0)]-\frac{1}{3\gamma}\ln[V(\phi_e)]
+\left(\frac{1}{3\gamma}-\frac14\right)\ln[\rho_{\rm reh}]
\,,
\end{equation}
where $\rho_{\rm reh}$ is the energy density at reheating, and
$\gamma-1$ represents the average equation of state during oscillations
of the inflaton. For $V\propto\phi^n$, $\gamma=2n/(n+2)$ \cite{Turner:1983he}.
In particular, for $\phi^2$ inflation $\gamma=1$ and the universe expands as
matter dominated during inflaton oscillations, whereas for $\phi^4$ inflation $\gamma=4/3$
and the universe expands as radiation dominated.
In the latter case $N_0$ does not depend on $\rho_{\rm reh}$. Note that with quantum corrections
included in the potential, $\gamma$ will in principle deviate from its tree level
value. However this effect is quite negligible since the tree level term dominates
at low values of $\phi$ where inflation has ended.

\begin{table*}[t]
{\centering
\caption{The inflationary parameters for the potential $V=(1/2)m^2\phi^2-\kappa\phi^4\ln(\phi/m_P)$ \\
(in units $m_P=1$)} \label{phi2table}
\resizebox{!}{3.7cm}{
\begin{tabular}{r@{\hspace{.5cm}}r@{\hspace{.5cm}}r@{\hspace{.5cm}}r@{\hspace{.5cm}}r@{\hspace{.5cm}}r@{\hspace{.5cm}}r@{\hspace{.5cm}}r@{\hspace{.5cm}}r@{\hspace{.5cm}}r}
\hline \hline
 $\log_{10}(\kappa)$ & $m\ (10^{-6})$ & $\phi_e$ & $\phi_0$ & $V(\phi_0)^{1/4}$
& $N_0$ & $u_0$ & $n_s$  & $r$ & $\alpha\ (10^{-4})$    \\
\hline \hline
\multicolumn{10}{c}{$V=(1/2)m^2\phi^2$ (assuming $\rho_{\rm reh}=10^{-16}m^2 m_P^2$)}\\
\hline
& 6.437 & 1.457 & 15.26 & 0.008334 & 58.31 & & 0.9657 & 0.1349 & -5.901\\
\hline \hline
\multicolumn{10}{c}{$\phi^2$ branch}\\\hline
-16 & 6.434 & 1.457 & 15.25 & 0.008322 & 58.31 & 319.4 & 0.9657 & 0.1341 & -5.901 \\\hline
-15 & 6.383 & 1.457 & 15.15 & 0.008204 & 58.47 & 30.2 & 0.9656 & 0.1267 & -5.853 \\\hline
-14.5 & 6.245 & 1.457 & 14.83 & 0.007891 & 58.5 & 8.355 & 0.9645 & 0.1085 & -5.647 \\\hline
-14.2 & 5.798 & 1.457 & 14.19 & 0.007212 & 58.43 & 3.165 & 0.9591 & 0.07567 & -4.423 \\\hline
-14.11 & 4.917 & 1.456 & 13.35 & 0.006241 & 58.23 & 2.067 & 0.9459 & 0.04239 & -1.254\\\hline
\hline
\multicolumn{10}{c}{Hilltop branch}\\
\hline
-14.11 & 4.917 & 1.456 & 13.35 & 0.006241 & 58.23 & 2.067 & 0.9459 & 0.04239 & -1.254\\\hline
-14.2 & 3.628 & 1.456 & 12.35 & 0.005019 & 57.93 & 0.3324 & 0.9219 & 0.01769 & 3.196 \\\hline
-14.5 & 2.146 & 1.455 & 11.18 & 0.003603 & 57.48 & 0.1447 & 0.8852 & 0.004665 & 6.022 \\\hline
-15 & 1.032 & 1.455 & 10.04 & 0.002344 & 56.88 & 0.0531 & 0.8424 & 0.000826 & 5.236 \\\hline
-16 & 0.268 & 1.453 & 8.617 & 0.001103 & 55.86 & 0.0102 & 0.7762 & 0.000039 & 2.254 \\\hline
\hline
\end{tabular} }
\par} \centering
\end{table*}

First, assume that $\lambda\ll m^2 / \phi^2$ during inflation, so that inflation is
primarily driven by the quadratic $\phi^2$ term.
For the tree level potential $V=(1/2)m^2\phi^2$, \eq{efold1} gives $N_0\simeq\phi_0^2/4$.
Using \eq{perturb}, $m\simeq1.6\times10^{13}$ GeV. Using the above definitions
we also obtain
\begin{eqnarray}
n_s \!&=&\! 1 - 8/\phi^2=1-2/N\,,\\
r \!&=&\! 32/\phi^2=8/N \,,\\
\alpha \!\!&=&\!\! -32/\phi^4=-2/N^2\,.
\end{eqnarray}
The number of e-folds is given by \eq{nuk}.
Assuming $m_N\ll m$, the inflaton decay rate $\Gamma_{\phi}=h^2m/(8\pi)$
(where $h^2<m$) and $\rho_{\rm reh}\simeq\kappa m^2 m_P^2$.

We can simplify the discussion of the potential with the loop correction
by treating $\ln\phi$ as constant. We then have
\begin{align}
V&=\frac12m^2\phi^2-\kappa\phi^4\ln\phi\,,\\
V^{\prime}&\simeq m^2\phi-4\kappa\phi^3\ln\phi\,,\\\label{perturb3}
\Delta_R&\simeq\frac{1}{4\sqrt3\pi}\sqrt\kappa\phi^3\frac{(u+1)^{3/2}}{u}\,,
\end{align}
where in \eq{perturb3} we have defined
\begin{equation}
u\equiv\frac{m^2}{2\kappa\phi^2\ln\phi}-2\,.
\end{equation}
The inflationary parameters are given by
\begin{eqnarray}
n_s \!&\simeq&\! 1 - \frac{8}{\phi^2}\left[\frac{u^2+(3/2)u+2}{(u+1)^2}\right]\,,\\
r \!&\simeq&\! \frac{32}{\phi^2}\left[\frac{u^2}{(u+1)^2}\right] \,,\\
\alpha \!\!&\simeq&\!\! -\frac{32}{\phi^4}\left[\frac{u(u^3+3u^2+2u-3)}{(1+u)^4}\right]\,.
\end{eqnarray}

\begin{figure}[t]
\includegraphics[height=.225\textheight]{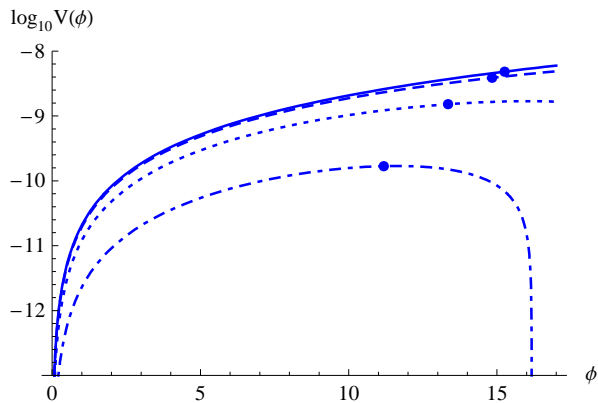}
\caption{The tree level potential (solid), the $\phi^2$ and hilltop
solution potentials for $\log_{10}(\kappa)=-14.5$ (dashed and dot-dashed),
and the potential for $\log_{10}(\kappa)=-14.11$ where the two solutions
meet (dotted). The points on the curves denote $\phi_0$.} \label{vfig}
\end{figure}

The numerical solutions are obtained (without the constant $\ln\phi$ approximation,
except for calculating $u_0$)
using \eq{perturb}, \eq{efold1} and \eq{nuk}.
(We also include the next to leading order corrections in the slow roll
expansion, see Appendix.)
One way to obtain the solutions is to fix $\kappa$ and scan over $m$
(with $\phi_0$ calculated for each $m$ value using \eq{perturb})
until $N_0$ matches \eq{nuk}. There are two solutions for a given value of
$\kappa$. From \eq{perturb3}, in the large $u_0$ limit ($u_0\gg1$ or
$m^2\gg4\kappa\phi_0^2\ln\phi_0$) a solution is obtained with $\kappa\propto1/u_0$.
In the small $u_0$ limit ($u_0\ll1$ or
$m^2\approx4\kappa\phi_0^2\ln\phi_0$), $\kappa\propto u_0^2$.
The two solutions meet at $u_0\sim1$, giving a maximum value of
$\kappa\sim(\sqrt6\pi\Delta_R)^2/\phi_0^6$. For larger values of $\kappa$,
it is not possible to satisfy the $\Delta_R$ and $N_0$ constraints
simultaneously, since the duration of inflation becomes too short
for the lower $\phi_0$ values required to keep $\Delta_R$ fixed.

\begin{figure}[t]
\psfrag{k}{\footnotesize{$\log_{10}(\kappa)$}}
\includegraphics[height=.225\textheight]{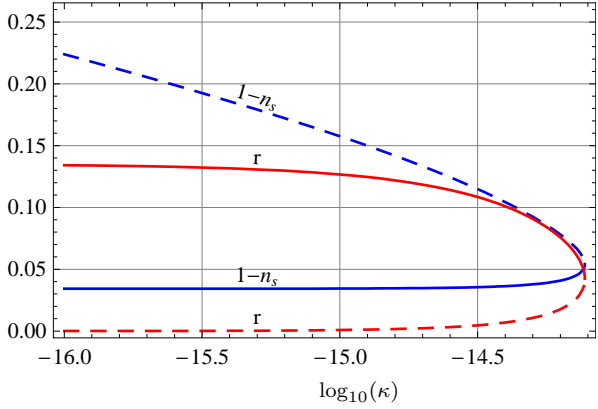}
\caption{$1-n_s$ and $r$ vs. $\kappa$ for the potential
$V=(1/2)m^2\phi^2-\kappa\phi^4\ln(\phi/m_P)$. Solid and
dashed curves correspond to $\phi^2$ and hilltop  branches
respectively.} \label{phi2fig}
\end{figure}

We call the large $u_0$ solution the $\phi^2$ solution, and the other
the hilltop solution \cite{Boubekeur:2005zm}. For the $\phi^2$ solution,
$u_0\to\infty$ as $\kappa\to0$. The predictions for $V=(1/2)m^2\phi^2$ are recovered
for $u_0\gg1$. On the other hand, for the hilltop  solution $u_0\to0$ as $\kappa\to0$.
With $u_0\ll1$, $n_s\approx1-16/\phi_0^2$ and $r$ is suppressed by $u^2$.
For the $\phi^2$ solution the local maximum of the potential and $\phi_0$ is at higher
values, whereas for the hilltop solution inflation occurs closer to the local maximum
(see \fig{vfig} and Table \ref{phi2table}). As the value of $\kappa$ is increased, the two branches of solutions
approach each other and they meet at $\kappa\simeq8\times10^{-15}$ (see \fig{phi2fig}).

Note that the one loop contribution to $\lambda$ is of order $(4!)\kappa$, which is $\sim m^2 / \phi^2$
in the parameter range where the $\kappa$ term has a significant effect on
inflationary observables. In this case our assumption $\lambda\ll m^2 / \phi^2$
corresponds to the renormalized coupling being small compared to the one loop contribution.

Alternatively, assume that $\lambda\gg m^2 / \phi^2$ during inflation, so that inflation is
primarily driven by the quartic term.
For the tree level potential $V=(1/4!)\lambda\phi^4$, \eq{efold1} gives $N_0\simeq\phi_0^2/8$.
Using \eq{perturb}, $\lambda\simeq8\times10^{-13}$. We also obtain
\begin{eqnarray}
n_s \!&=&\! 1 - 24/\phi^2=1-3/N\,,\\
r \!&=&\! 128/\phi^2=16/N \,,\\
\alpha \!\!&=&\!\! -192/\phi^4=-3/N^2\,.
\end{eqnarray}

\begin{table*}[t]
{\centering
\caption{The inflationary parameters for the potential $V=(1/4!)\lambda\phi^4-\kappa\phi^4\ln(\phi/m_P)$ \\
(in units $m_P=1$)} \label{phi4table}
\resizebox{!}{3.7cm}{
\begin{tabular}{r@{\hspace{.5cm}}r@{\hspace{.5cm}}r@{\hspace{.5cm}}r@{\hspace{.5cm}}r@{\hspace{.5cm}}r@{\hspace{.5cm}}r@{\hspace{.5cm}}r@{\hspace{.5cm}}r@{\hspace{.5cm}}r}
\hline \hline
 $\log_{10}(\kappa)$ & $\log_{10}(\lambda)$ & $\phi_e$ & $\phi_0$ & $V(\phi_0)^{1/4}$
& $N_0$ & $v_0$ & $n_s$  & $r$ & $\alpha\ (10^{-4})$    \\
\hline \hline
\multicolumn{10}{c}{$V=(1/4!)\lambda\phi^4$}\\\hline
& -12.07 & 2.53 & 22.39 & 0.009737 & 62.55 & & 0.9517 & 0.251 & -7.637\\\hline
\hline
\multicolumn{10}{c}{$\phi^4$ branch}\\\hline
-15. & -12.03 & 2.516 & 22.31 & 0.00972 & 62.54 & 143.1 & 0.9519 & 0.2493 & -7.606\\\hline
-14. & -11.78 & 2.438 & 21.69 & 0.009558 & 62.43 & 14.08 & 0.9539 & 0.2331 & -7.372\\\hline
-13.5 & -11.49 & 2.369 & 20.49 & 0.009058 & 62.2 & 3.834 & 0.9575 & 0.1881 & -7.025\\\hline
-13.3 & -11.36 & 2.338 & 19.35 & 0.008344 & 61.97 & 1.762 & 0.9577 & 0.1355 & -6.261\\\hline
-13.24 & -11.33 & 2.319 & 18.23 & 0.007421 & 61.74 & 0.9184 & 0.9512 & 0.08476 & -3.725\\\hline
\hline
\multicolumn{10}{c}{Hilltop branch}\\\hline
-13.24 & -11.33 & 2.319 & 18.23 & 0.007421 & 61.74 & 0.9184 & 0.9512 & 0.08476 & -3.725\\\hline
-13.3 & -11.41 & 2.305 & 17.11 & 0.006329 & 61.49 & 0.4937 & 0.9359 & 0.04481 & 0.9321\\\hline
-13.5 & -11.63 & 2.292 & 15.85 & 0.004985 & 61.14 & 0.2391 & 0.9088 & 0.01718 & 6.326\\\hline
-14. & -12.15 & 2.276 & 14.15 & 0.003225 & 60.57 & 0.0799 & 0.8618 & 0.002978 & 8.232\\\hline
-15. & -13.18 & 2.256 & 12.15 & 0.001534 & 59.69 & 0.0151 & 0.7959 & 0.000149 & 4.078\\\hline
\hline
\end{tabular} }
\par} \centering
\end{table*}

Including the loop correction we have
\begin{align}
V&=\frac{\phi^4}{24}(\lambda-24\kappa\ln\phi)\,,\\
V^{\prime}&=\frac{\phi^3}{6}(\lambda-6\kappa-24\kappa\ln\phi)\,,\\\label{perturb2}
\Delta_R&\simeq\frac{\sqrt3}{48\pi}\sqrt\kappa\phi^3\frac{(v+1)^{3/2}}{v}\,,
\end{align}
where in \eq{perturb2} we have defined
\begin{equation}
v\equiv\frac1{6\kappa}(\lambda-6\kappa-24\kappa\ln\phi)\,.
\end{equation}
The inflationary parameters are given by
\begin{eqnarray}
n_s \!&=&\! 1 - \frac{24}{\phi^2}\left[\frac{v^2+v/3+4/3}{(v+1)^2}\right]\,,\\
r \!&=&\! \frac{128}{\phi^2}\left[\frac{v^2}{(v+1)^2}\right] \,,\\
\alpha \!\!&=&\!\! -\frac{192}{\phi^4}\left[\frac{v(v^3+(4/3)v^2+5v-10/3)}{(1+v)^4}\right]\,.
\end{eqnarray}

\begin{figure}[t]
\psfrag{k}{\footnotesize{$\log_{10}(\kappa)$}}
\includegraphics[height=.225\textheight]{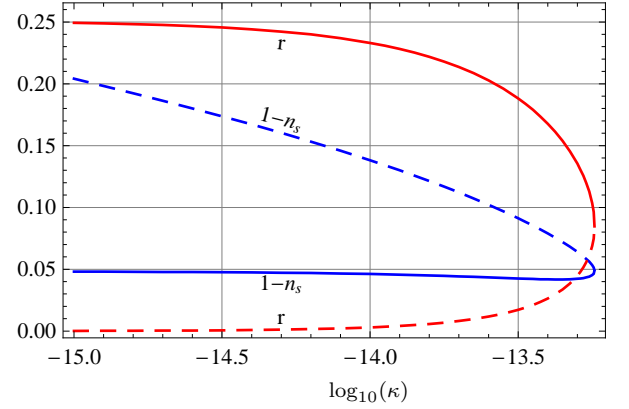}
\caption{$1-n_s$ and $r$ vs. $\kappa$ for the potential
$V=(1/4!)\lambda\phi^4-\kappa\phi^4\ln(\phi/m_P)$. Solid and
dashed curves correspond to $\phi^4$ and hilltop  branches
respectively.}\label{phi4fig}
\end{figure}

The numerical results are displayed in \fig{phi4fig} and Table \ref{phi4table}.
As before, there are two solutions for a given value of
$\kappa$. We call the large $v_0$ solution the $\phi^4$ solution, and the other
the hilltop solution. The predictions for $V=(1/4!)\lambda\phi^4$ are recovered
for $v_0\gg1$, or $\lambda\gg24\kappa\ln\phi_0$. Since $\phi_0^2=8N_0$ for $\phi^4$
potential, this corresponds to $\lambda\gg75\kappa$.
As the value of $\kappa$ is increased, the two branches of solutions
approach each other and they meet at
$\kappa\simeq(4\sqrt6\pi\Delta_R)^2/\phi_0^6\simeq6\times10^{-14}$.

\begin{figure}[t]
\includegraphics[height=.225\textheight]{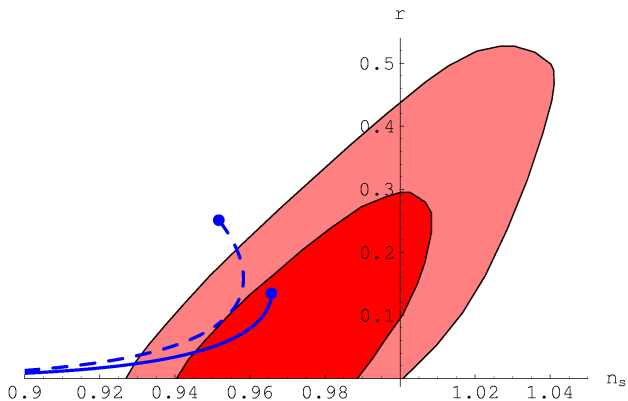}
\caption{Tensor to scalar ratio $r$ vs. the spectral index $n_s$ for the potential
$V=(1/2)m^2\phi^2-\kappa\phi^4\ln(\phi/m_P)$
 (solid curve) and for the potential $V=(1/4!)\lambda\phi^4-\kappa\phi^4\ln(\phi/m_P)$
(dashed curve).
The WMAP contours (68\% and 95\% CL) are taken from \ocite{Dunkley:2008ie}.
The points on the curves correspond to the tree level
predictions for $\phi^2$ and $\phi^4$ potentials.}\label{wmapfig}
\end{figure}

To summarize, in this paper we have considered the impact radiative
corrections can have on  chaotic inflation predictions with $\phi^2$
and $\phi^4$ potentials. A Yukawa coupling of $\phi$, in particular,
induces corrections to the inflationary potential with a negative
sign, which can lower $r$. We display the possible range of values
for the inflationary parameters including such corrections. As shown
in \fig{wmapfig}, although $\phi^4$ inflation seems excluded at tree
level, it can become compatible with WMAP when this correction is
included. The current WMAP limits imply $r\gtrsim0.02$
($r\gtrsim0.04$) for the $\phi^2$ ($\phi^4$) model, which therefore
suggests that signatures of primordial gravitational waves should be
observed in the near future.

Finally we note that radiative corrections can also significantly alter the
inflationary predictions of other models. For instance, the Yukawa coupling
induced correction considered here can lead to a red-tilted spectrum (including
$n_s\approx0.96$ as favored by WMAP) in the non-supersymmetric hybrid inflation
model \cite{Linde:1993cn}, which otherwise predicts a blue spectrum. 

\begin{acknowledgments}
This work is partially supported by the US DOE under contract number
DE-FG02-91ER40626 (Q.S.).
\end{acknowledgments}

\section*{Appendix}

We provide here the next to leading order formulae for calculating $n_s$ and $r$
that we have used \cite{Stewart:1993bc}:
\begin{align}
\begin{split}
\Delta_R&=\frac{1}{2\sqrt{3}\pi }\frac{V^{3/2}}{|V^{\prime}|}\bigg[ 1- \left(3C+\frac16\right)\epsilon
 \\& \quad+\left(C-\frac13 \right) \eta\bigg]\,,
\end{split}\\
\begin{split}
n_s& =1+2\bigg[ -3\epsilon + \eta - \left(\frac53 + 12 C\right)\epsilon^2
\\& \quad+ \left(8C-1\right) \epsilon\eta +\frac13 \eta^2
-\left(C-\frac13\right) \xi^2 \bigg] \,,
\end{split}\\
r&=16\epsilon\left[1+\frac23(3C-1)(2\epsilon-\eta)\right]\,,
\end{align}
where $C=\ln2+\gamma_E-2\approx-0.7296$. Inflation ends at $\epsilon_H=1$
and
\begin{align}
N_0&=\int^{\phi_0}_{\phi_e}\frac{\rm{d}\phi}{\sqrt{2\epsilon_H}}\,,\\ 
\begin{split}
\epsilon_H&=2\left(\frac{H^{\prime}(\phi)}{H(\phi)}\right)^2
=\epsilon\bigg(1-\frac43\epsilon+\frac23\eta+\frac{32}{9}\epsilon^2
  \\& \hspace{2.6cm}
+\frac59\eta^2-\frac{10}{3}\epsilon\eta+\frac29\xi^2+\ldots\bigg)\,.
\end{split}
\end{align}


\begin{thebibliography}{11}
\expandafter\ifx\csname natexlab\endcsname\relax\def\natexlab#1{#1}\fi
\expandafter\ifx\csname bibnamefont\endcsname\relax
  \def\bibnamefont#1{#1}\fi
\expandafter\ifx\csname bibfnamefont\endcsname\relax
  \def\bibfnamefont#1{#1}\fi
\expandafter\ifx\csname citenamefont\endcsname\relax
  \def\citenamefont#1{#1}\fi
\expandafter\ifx\csname url\endcsname\relax
  \def\url#1{\texttt{#1}}\fi
\expandafter\ifx\csname urlprefix\endcsname\relax\def\urlprefix{URL }\fi
\providecommand{\bibinfo}[2]{#2}
\providecommand{\eprint}[2][]{\url{#2}}

\bibitem[{\citenamefont{Linde}(1983)}]{Linde:1983gd}
\bibinfo{author}{\bibfnamefont{A.~D.} \bibnamefont{Linde}},
  \bibinfo{journal}{Phys. Lett.} \textbf{\bibinfo{volume}{B129}},
  \bibinfo{pages}{177} (\bibinfo{year}{1983}).

\bibitem[{\citenamefont{Dunkley et~al.}(2008)}]{Dunkley:2008ie}
\bibinfo{author}{\bibfnamefont{J.}~\bibnamefont{Dunkley}} \bibnamefont{et~al.}
  (\bibinfo{collaboration}{WMAP}), \eprint{arXiv:0803.0586}.

\bibitem[{\citenamefont{Komatsu et~al.}(2008)}]{Komatsu:2008hk}
\bibinfo{author}{\bibfnamefont{E.}~\bibnamefont{Komatsu}} \bibnamefont{et~al.}
  (\bibinfo{collaboration}{WMAP}), \eprint{arXiv:0803.0547}.

\bibitem[{\citenamefont{Coleman and Weinberg}(1973)}]{Coleman:1973jx}
\bibinfo{author}{\bibfnamefont{S.~R.} \bibnamefont{Coleman}} \bibnamefont{and}
  \bibinfo{author}{\bibfnamefont{E.}~\bibnamefont{Weinberg}},
  \bibinfo{journal}{Phys. Rev.} \textbf{\bibinfo{volume}{D7}},
  \bibinfo{pages}{1888} (\bibinfo{year}{1973}).

\bibitem[{\citenamefont{Shafi and {\c
  S}eno$\breve{\textrm{g}}$uz}(2006)}]{Shafi:2006cs}
\bibinfo{author}{\bibfnamefont{Q.}~\bibnamefont{Shafi}} \bibnamefont{and}
  \bibinfo{author}{\bibfnamefont{V.~N.} \bibnamefont{{\c
  S}eno$\breve{\textrm{g}}$uz}}, \bibinfo{journal}{Phys. Rev.}
  \textbf{\bibinfo{volume}{D73}}, \bibinfo{pages}{127301}
  (\bibinfo{year}{2006}) \eprint{[arXiv:astro-ph/0603830]}.

\bibitem{Kawasaki:2000yn}
  M.~Kawasaki, M.~Yamaguchi and T.~Yanagida,
  Phys.\ Rev.\ Lett.\  {\bf 85}, 3572 (2000)
  [arXiv:hep-ph/0004243].

\bibitem{Linde:2005ht}
  A.~D.~Linde,
  {\em Particle Physics and Inflationary Cosmology} (Chur, Switzerland: Harwood, 1990)
  [arXiv:hep-th/0503203].

\bibitem{Linde:2007fr}
  A.~D.~Linde,
  Lect.\ Notes Phys.\  {\bf 738}, 1 (2008)
  [arXiv:0705.0164].

\bibitem{Cirigliano:2004yh}
  D.~Cirigliano, H.~J.~de Vega and N.~G.~Sanchez,
  Phys.\ Rev.\   {\bf D71}, 103518 (2005)
  [arXiv:astro-ph/0412634].
Also see
  R.~Kallosh and A.~Linde,
  JCAP {\bf 0704}, 017 (2007)
  [arXiv:0704.0647] and references therein.

\bibitem[{\citenamefont{Miao and Woodard}(2006)}]{Miao:2006pn}
\bibinfo{author}{\bibfnamefont{S.-P.} \bibnamefont{Miao}} \bibnamefont{and}
  \bibinfo{author}{\bibfnamefont{R.~P.} \bibnamefont{Woodard}},
  \bibinfo{journal}{Phys. Rev.} \textbf{\bibinfo{volume}{D74}},
  \bibinfo{pages}{044019} (\bibinfo{year}{2006}) \eprint{[arXiv:gr-qc/0602110]}.

\bibitem{Sloth:2006az}
  M.~S.~Sloth,
  Nucl.\ Phys.\   {\bf B748}, 149 (2006) [arXiv:astro-ph/0604488];
  Nucl.\ Phys.\   {\bf B775}, 78 (2007) [arXiv:hep-th/0612138].

\bibitem{Vilenkin:1983xq}
  A.~Vilenkin,
  Phys.\ Rev.\  {\bf D27}, 2848 (1983).

\bibitem[{\citenamefont{Boubekeur and Lyth}(2005)}]{Boubekeur:2005zm}
\bibinfo{author}{\bibfnamefont{L.}~\bibnamefont{Boubekeur}} \bibnamefont{and}
  \bibinfo{author}{\bibfnamefont{D.~H.} \bibnamefont{Lyth}},
  \bibinfo{journal}{JCAP} \textbf{\bibinfo{volume}{0507}}, \bibinfo{pages}{010}
  (\bibinfo{year}{2005}) \eprint{[arXiv:hep-ph/0502047]}.

\bibitem[{\citenamefont{Liddle and Lyth}(2000)}]{Liddle:2000cg}
\bibinfo{author}{\bibfnamefont{A.~R.} \bibnamefont{Liddle}} \bibnamefont{and}
  \bibinfo{author}{\bibfnamefont{D.~H.} \bibnamefont{Lyth}},
  \emph{\bibinfo{title}{Cosmological inflation and large-scale structure}}
  (\bibinfo{publisher}{Cambridge, UK: Univ. Pr.}, \bibinfo{year}{2000}).

\bibitem[{\citenamefont{Kolb and Turner}(1990)}]{Kolb:1990vq}
\bibinfo{author}{\bibfnamefont{E.~W.} \bibnamefont{Kolb}} \bibnamefont{and}
  \bibinfo{author}{\bibfnamefont{M.~S.} \bibnamefont{Turner}},
  \emph{\bibinfo{title}{The Early universe}}
  (\bibinfo{publisher}{Addison-Wesley}, \bibinfo{year}{1990});
\bibinfo{author}{\bibfnamefont{A.~R.} \bibnamefont{Liddle}} \bibnamefont{and}
  \bibinfo{author}{\bibfnamefont{S.~M.} \bibnamefont{Leach}},
  \bibinfo{journal}{Phys. Rev.} \textbf{\bibinfo{volume}{D68}},
  \bibinfo{pages}{103503} (\bibinfo{year}{2003}) \eprint{[arXiv:astro-ph/0305263]}.

\bibitem[{\citenamefont{Turner}(1983)}]{Turner:1983he}
\bibinfo{author}{\bibfnamefont{M.~S.} \bibnamefont{Turner}},
  \bibinfo{journal}{Phys. Rev.} \textbf{\bibinfo{volume}{D28}},
  \bibinfo{pages}{1243} (\bibinfo{year}{1983}).

\bibitem{Linde:1993cn}
  A.~D.~Linde,
  Phys.\ Rev.\   {\bf D49}, 748 (1994)
  [arXiv:astro-ph/9307002].

\bibitem[{\citenamefont{Stewart and Lyth}(1993)}]{Stewart:1993bc}
\bibinfo{author}{\bibfnamefont{E.~D.} \bibnamefont{Stewart}} \bibnamefont{and}
  \bibinfo{author}{\bibfnamefont{D.~H.} \bibnamefont{Lyth}},
  \bibinfo{journal}{Phys. Lett.} \textbf{\bibinfo{volume}{B302}},
  \bibinfo{pages}{171} (\bibinfo{year}{1993}) \eprint{[arXiv:gr-qc/9302019]};
\bibinfo{author}{\bibfnamefont{E.~W.} \bibnamefont{Kolb}} \bibnamefont{and}
  \bibinfo{author}{\bibfnamefont{S.~L.} \bibnamefont{Vadas}},
  \bibinfo{journal}{Phys. Rev.} \textbf{\bibinfo{volume}{D50}},
  \bibinfo{pages}{2479} (\bibinfo{year}{1994}) \eprint{[arXiv:astro-ph/9403001]};
\bibinfo{author}{\bibfnamefont{A.~R.} \bibnamefont{Liddle}},
  \bibinfo{author}{\bibfnamefont{P.}~\bibnamefont{Parsons}}, \bibnamefont{and}
  \bibinfo{author}{\bibfnamefont{J.~D.} \bibnamefont{Barrow}},
  \bibinfo{journal}{Phys. Rev.} \textbf{\bibinfo{volume}{D50}},
  \bibinfo{pages}{7222} (\bibinfo{year}{1994}) \eprint{[arXiv:astro-ph/9408015]}.

\end{thebibliography}
\end{document}